\documentclass[aps,prl,amsmath,amssymb,reprint,superscriptaddress,]{revtex4-1}

\usepackage{physics}
\usepackage{graphicx}
\usepackage{grffile}
\usepackage{float}
\usepackage{color,soul}
\usepackage{hyperref}
\usepackage{xspace}
\usepackage{makecell}

\newcommand{\NBNV}{\ensuremath{\text{N}_\text{B}\text{V}_\text{N}}\xspace}
\newcommand{\CBNV}{\ensuremath{\text{C}_\text{B}\text{V}_\text{N}}\xspace}
\newcommand{\NV}{\ensuremath{\text{V}_\text{N}}\xspace}
\newcommand{\CB}{\ensuremath{\text{C}_\text{B}}\xspace}
\newcommand{\Epristine}{{\ensuremath{E_{\text{pst}}}}\xspace}
\newcommand{\EQP}{{\ensuremath{E_{\text{QP}}}}\xspace}
\newcommand{\Erlx}{{\ensuremath{E_{\text{rlx}}}}\xspace}
\newcommand{\Angst}{\AA\xspace}

\newcommand{\comment}[1]{}

\begin{document}

\title[]{First-principles Engineering of Charged Defects\\for Two-dimensional Quantum Technologies}

\author{Feng Wu}
\affiliation{The Department of Chemistry and Biochemistry, University of California, Santa Cruz, 95064 CA, United States}

\author{Andrew Galatas}
\affiliation{The Department of Physics, University of California, Santa Cruz, 95064 CA, United States}

\author{Ravishankar Sundararaman}
\affiliation{The Department of Materials Science and Engineering, Rensselaer Polytechnic Institute, 110 8th street, Troy, New York 12180, USA}

\author{Dario Rocca}
\affiliation{Universit\'e de Lorraine, CRM2, UMR 7036, 54506 Vandoeuvre-l\`es-Nancy, France}
\affiliation{CNRS, CRM2, UMR 7036, 54506 Vandoeuvre-l\`es-Nancy, France}

\author{Yuan Ping}
\email{yuanping@ucsc.edu}
\affiliation{The Department of Chemistry and Biochemistry, University of California, Santa Cruz, 95064 CA, United States}


\begin{abstract}
Charged defects in 2D materials have emerging applications in quantum technologies
such as quantum emitters and quantum computation.
Advancement of these technologies requires rational design of ideal defect centers,
demanding reliable computation methods for quantitatively accurate prediction of defect properties.
We present an accurate, parameter-free and efficient procedure to evaluate quasiparticle
defect states and thermodynamic charge transition levels of defects in 2D materials.
Importantly, we solve critical issues that stem from the strongly anisotropic screening in 2D materials,
that have so far precluded accurate prediction of charge transition levels in these materials.
Using this procedure, we investigate various defects in monolayer hexagonal boron nitride (h-BN)
for their charge transition levels, stable spin states and optical excitations. We identify \CBNV (nitrogen vacancy adjacent to carbon substitution of boron)
to be the most promising defect candidate for scalable quantum bit and emitter applications.
\end{abstract}


\maketitle

Two-dimensional (2D) materials such as graphene, hexagonal Boron Nitride (h-BN)
and transition metal dichalcogenides exhibit a wide range of remarkable properties
at atomic-scale layer thicknesses,
holds promise for both conventional and
new optoelectronic functionality at drastically reduced dimensions~\cite{Butler2013,tianshu1,Novoselov2012,Pakdel2013,Zhangting2017}.
It is well established that
point defects play a central role in the properties of bulk 3D semiconductors
 but their corresponding role in 2D materials is not yet well understood.
In particular, the weak screening environment 
surrounding the defect charge distribution
and the strong confinement of wavefunctions due to the atomic-scale thickness
could lead to vastly different behavior compared to conventional semiconductors.

Defects in 2D materials such as h-BN show promise as polarized
and ultra-bright single-photon emitters at room temperature,
with potentially better scalability than the long-studied
nitrogen-vacancy center(NV) in diamond\cite{Weber2010,Koehl2015,hosung2016}
for emerging applications in nanophotonics and quantum information.\cite{Jelezko2004}
Progress beyond initial experimental demonstration of promising properties requires
rational design and development of quantum defects in 2D materials that exhibit
high emission rate, long coherence time, single photon purity and stability.
Specifically, the promising defects should have the following properties: defect levels should be deep (far from band edges) to avoid resonance with
the bulk band edges and thereby exhibit long coherence time;\cite{Tran2016,Shotan2016}
optically-addressable spin conserving excitations facilitate exploiting spin-selective
decays in high-spin defect states, similar to the NV center in diamond;\cite{Exarhos2016,Choi2016,Tawfik2017}
anisotropic polarization of the defect states in combination
with quantum bits could provide a pathway to quantum optical computation.

\begin{figure}
\includegraphics[width=\columnwidth]{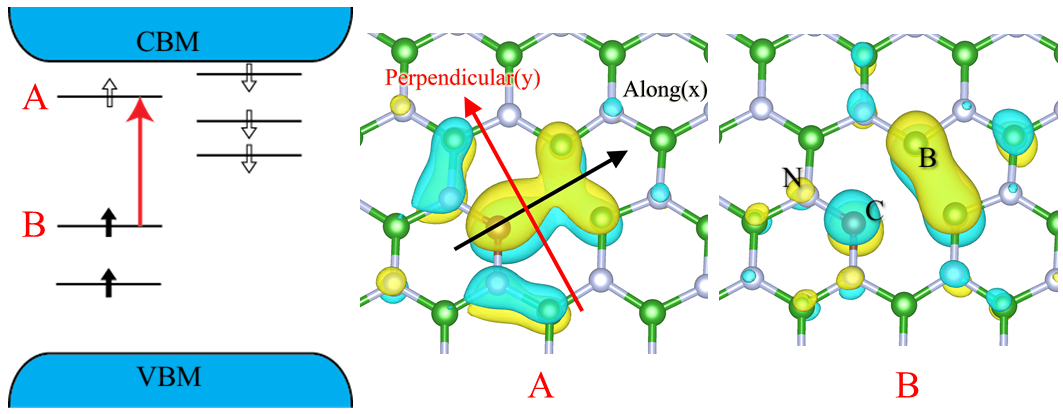}
\caption{Left: \CBNV defect energy levels in monolayer BN with spin up (up arrow) and spin down (down arrow) channels respectively.
The black filled arrows represent occupied states and unfilled arrows represent unoccupied states. The red arrow represents the bright transition
between two defect states. Right: The wavefunctions for the two defect states (``A'' and ``B'') that have the bright optical transition. ``Perpendicular'' and ``Along''
are two orthogonal directions in the plane; only the ``Perpendicular'' direction has the bright transition.}
\label{fig:opt-CBNV}
\end{figure}

\comment{
\begin{figure}
\begin{minipage}{0.45\columnwidth}
\includegraphics[width=\columnwidth]{./CBNV.9.notated.png}
\end{minipage}\quad%
\begin{minipage}{0.45\columnwidth}
\includegraphics[width=\columnwidth]{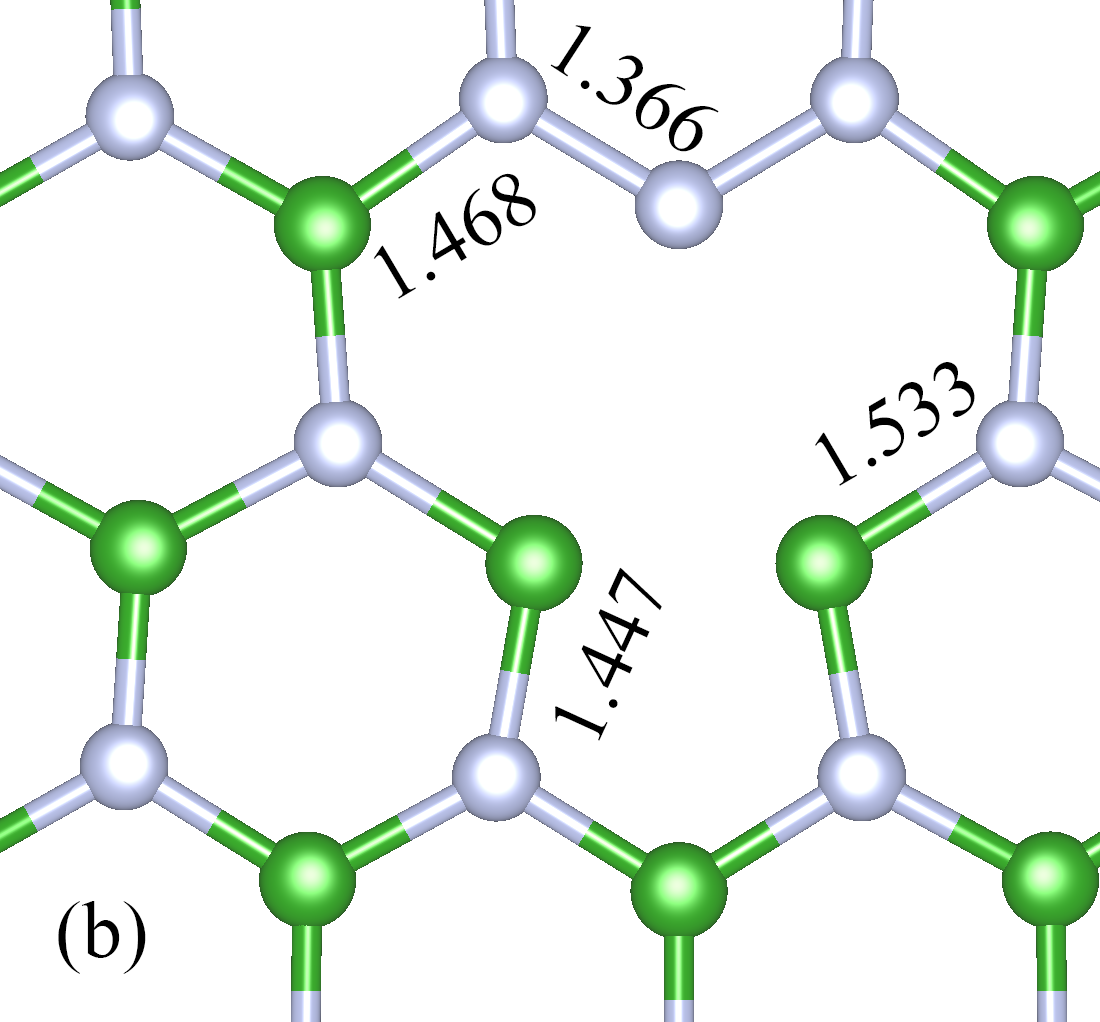}
\end{minipage}
\caption{Defects in hexagonal Boron Nitride (h-BN) with promise for
quantum emitter applications: nitrogen vacancy adjacent to boron
substituted either by (a) carbon `\CBNV' or by (b) nitrogen `\NBNV'.
Bond lengths (shown in \Angst units) can be compared to a
B$-$N bond length of 1.442~\Angst in pristine h-BN.
\label{fig:geometry}}
\end{figure}
}

In this work, we use first-principles methods to theoretically
investigate the suitability of several complex defects in
monolayer h-BN for quantum bit and emitter applications.
We choose n-type defects which are closely related to common intrinsic and extrinsic defects in BN
and can potentially create several occupied defect levels
in the band gap and high spin states.
For each candidate defect, we examine the charge transition level (CTL)
which determines the stability ranges of various charge states of each defect.
For each stable charge state, we evaluate spin states and
optical excitations along different polarization directions.
With these calculations, we will show \CBNV (a nitrogen vacancy adjacent to boron substituted by carbon as shown in Fig.~\ref{fig:opt-CBNV}) to be the most promising
defect in 2D h-BN, analogous to the NV center in 3D diamond, which has a stable triplet ground state and a bright anisotropic optical transition between defect levels.

However, calculating properties such as CTLs and optical excitations of
charged defects in 2D materials present serious challenges for state-of-the-art
first-principles methods, which have so far limited the accuracy of previous calculations.
We start this Letter by outlining these challenges and then discuss
our methodology to address them.

The formation energy of a defect in charge state $q$, ionic coordinates $\textbf{R}$
and electron chemical potential $\varepsilon_F$ (often set to the valence band maximum
for insulators or semiconductors) is given by\cite{Freysoldt2014}
\begin{equation}
E^{f}_q(\textbf{R})[\varepsilon_F]=E_{q}(\textbf{R})-\Epristine+\sum_{i}\mu_{i}\Delta N_{i}+q\varepsilon_F.
\label{Ef}
\end{equation}
Here $E_q(\textbf{R})$ is the total energy of the system with the charged defect,
and $\Epristine$ is the total energy of the pristine system.
The third term on the right side accounts for the change $\Delta N_i$ in
number of atoms of element $i$ between these two configurations, with $\mu_i$
being the atomic chemical potential of that element in its stable form.
The charge transition level (CTL) is the value of the electron chemical potential
at which the stable charge state of the defect changes from $q$ to $q+1$,
which corresponds to equal formation energies of the $q$ and $q+1$ states,
and is therefore given by
\begin{equation}
\varepsilon^{q+1|q}=E^{f}_{q}(\textbf{R}_{q})-E^{f}_{q+1}(\textbf{R}_{q+1}),
\label{CTL}
\end{equation}
where $\textbf{R}_q$ are the ionic coordinates of the charge state $q$.

Within density-functional theory (DFT), CTLs can be determined
by calculating formation energies in Eq.~\ref{CTL} in their
respective equilibrium geometries, but this introduces two problems.
First, DFT calculations of defects employ periodic boundary conditions
on a supercell; formation energies of charged defects converge very
slowly with supercell size due to periodic charge interactions
and this is even more problematic for 2D materials.
Second, the well-known band gap problem and self-interaction errors within
standard DFT methods introduce significant errors in calculated CTLs,
even if the supercell convergence issue could be dealt with.

The second issue above can be effectively solved by combining
DFT with the many-body perturbation theory GW method~\cite{Govoni2015,Ping2013,Jain2011,Rinke2009}.
This involves rewriting the CTL calculation as\cite{Malashevich2014,Chen2015,Rinke2009}
\begin{multline}
\varepsilon^{q+1|q} = \underbrace{E^{f}_{q}(\textbf{R}_{q})-E^{f}_{q+1}(\textbf{R}_{q})}_\EQP
	+ \underbrace{E^{f}_{q+1}(\textbf{R}_{q})-E^{f}_{q+1}(\textbf{R}_{q+1})}_\Erlx,
\label{DFT+GW}
\end{multline}
by adding and subtracting $E^{f}_{q+1}(\textbf{R}_{q})$
(we note that the results are insensitive to the choice of path for defects in monolayer BN as discussed in SI).
The second pair of terms on the right-hand side of Eq.~\ref{DFT+GW}
is the structural relaxation energy \Erlx at the charge state $q$,
which can be calculated with reasonable accuracy at the DFT level
(provided we solve the periodic charge interaction issue).
The first pair of terms in Eq.~\ref{DFT+GW} is the quasiparticle (QP)
excitation energy \EQP at the fixed geometry $\textbf{R}_q$,
which can be calculated accurately using the GW method~\cite{Chen2017,Pham2013,Govoni2015}.
However, GW calculations of quasiparticle energies in 2D materials exhibit serious
convergence difficulties\cite{Qiu2016,Huser2013,Rasmussen2016} that make the calculations of charged defects
that require large supercells extremely challenging.

At this stage, Eq.~\ref{DFT+GW} provides accurate CTLs in principle,
provided we can address the periodic charge interaction issue
in formation energies of charged defects at the DFT level,
and resolve convergence issues for GW calculations of 2D materials.
Below, we discuss each of these two issues and our methodology to overcome them.


\emph{First}, the basic problem in charged defect formation energy calculations in DFT is
the spurious interaction of the charged defect with its periodic images and with the
uniform compensating background charge (necessary to make the total energy finite).
For 3D systems, correction schemes\cite{Freysoldt2009,Komsa2012,Kumagai2014,Wang2015a}
by removing the spurious periodic interaction from the DFT results using a model charge distribution for the defect
and a model dielectric response for the bulk material work
reliably well, because the self-energies of a model charge distribution both
with periodic boundary conditions and without i.e. the isolated case can be computed easily\cite{Freysoldt2009}.

\begin{figure}
\includegraphics[width=\columnwidth]{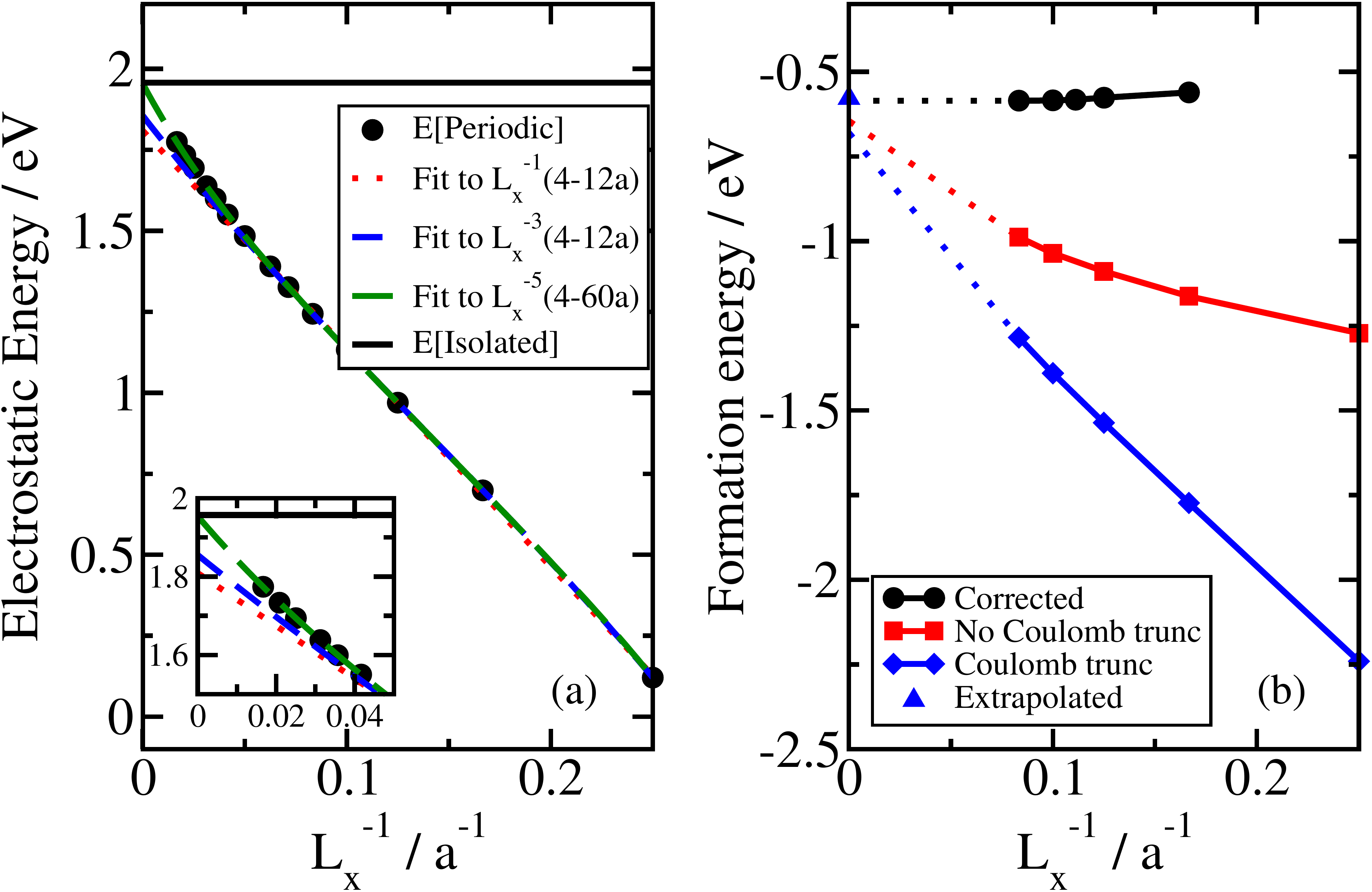}
\caption{Electrostatic self-energies of a model charge in a 2D slab with the periodic boundary condition (black dots) and the isolated boundary condition (black line).
Dashed red, blue and green lines are fitting curves to the periodic electrostatic self-energies with different order of polynomials. (b) Formation energies of \CB(+1) defect at different supercell
sizes $L$ (where $a = 2.50$~\Angst is the lattice constant of h-BN).
With conventional periodic Coulomb interactions (red squares), the cell size scales with $L_z = L_x$,
while with truncated Coulomb potentials (blue diamonds), $L_z$ is constant with a 15 bohr vacuum.
With our correction scheme (black dots), results are converged to within 10~meV in a 6$\times$6 a$^2$ cell.
The extrapolated result (blue triangle) includes a correction based on the fitted results from (a).
 \label{fig:model-charge-emodel-fit}}
\centering
\end{figure}

However, for 2D materials, the dielectric screening is strongly anisotropic
and localized to one atomic layer;
correction schemes now require a spatially-dependent anisotropic dielectric function,
whose spatial profile is not unambiguously defined.
Most importantly, calculation of the isolated charge self-energy for the correction
has so far relied on extrapolating periodic calculations in various supercell
sizes,\cite{Komsa2014} an approach we find here to be problematic
due to its strong nonlinear dependence on the supercell sizes (as shown in Fig.~\ref{fig:model-charge-emodel-fit}(a)).
This nonlinearity comes both from the highly anisotropic screening in monolayer 2D materials
and the spatial distribution of bound charge in the dielectric surrounding the model charge,
which has also been found in defects of MoS$_2$\cite{Noh2014}.

\comment{Fig.~\ref{fig:model-charge-emodel-fit}(a) illustrates the problem in
extrapolating the periodic self-energy of a (Gaussian) model charge distribution
(with the dielectric function of pristine monolayer BN) to the infinite limit.
For small system sizes ($L^{-1}>0.05 a^{-1}$, where $a = 2.50$~\Angst is the lattice constant of h-BN),
the variation appears
to be linear and extrapolation using these points alone will appear
to converge at low order (say including terms up to $L^{-3}$).
However, upon including larger cells ($L^{-1}<0.05 a^{-1}$),
a strong nonlinearity becomes apparent in the variation and
we need to include terms up to $L^{-5}$ for a reasonable fit.
Using a linear or cubic fit for extrapolation to the isolated charge energy
(as Ref.\citenum{Komsa2014} did) introduces an error $\sim 0.1 - 0.2$~eV for h-BN;
this could be even larger for high-dielectric systems such as MoS$_2$
This nonlinearity comes both from the highly anisotropic screening in monolayer 2D materials
and the spatial distribution of bound charge in the dielectric surrounding the model charge.}

We recently developed a robust scheme for calculating formation energies
of charged defects in bulk and interfaces,\cite{Sundararaman2017} that
(a) redefines DFT electrostatic potentials to avoid strong oscillations
near atom centers improving supercell convergence with geometry optimization,
(b) unambiguously defines a spatial dielectric profile
$\epsilon^{-1}(z)=\frac{-\partial\Delta V(z)}{E_0\partial z}$
as the change in the now-smooth total potential $\Delta V(z)$
upon applying a normal electric field $E_0$, and
importantly, (c) it also completely avoids the problematic extrapolation between supercell
sizes\cite{Komsa2014} (or convergence issues in image charge methods\cite{Komsa2012})
by using a spectral expansion in cylindrical Bessel functions for the
isolated electrostatic energy.
In this work, we extend all aspects of that approach to handle the anisotropic dielectric response
in 2D materials (see SI), including an exact calculation of the isolated electrostatic self-energy.
Fig.~\ref{fig:model-charge-emodel-fit}(a) shows that the conventional extrapolation
techniques line up to this result, but only when those fits are done to a high
enough order (e.g. fifth order). 

Fig.~\ref{fig:model-charge-emodel-fit}(b) shows that our charge correction scheme
(black dots) converges the \CB (+1) charged defect formation energy within 10 meV
in a 6$\times$6 supercell, with a converged value of -0.59~eV.
The formation energy without the correction and with isotropic supercell extrapolation
by a third order polynomial (red line) gives a similar result with Ref.\citenum{Komsa2014};
but fail to account for the nonlinearity of the periodic model charge self-energy with supercell sizes.
The difference between our new method (black dots) and this extrapolated result (red line)
in Fig.~\ref{fig:model-charge-emodel-fit}(b) is 0.12~eV,
which lines up exactly with the difference between third and fifth
order extrapolation in Fig.~\ref{fig:model-charge-emodel-fit}(a).
We therefore expect that previous predictions of charged defect formation energies
in 2D materials could routinely contain inaccuracies of this magnitude.


The \emph{second} major issue is the extremely slow numerical convergence of
the GW method for 2D materials,in part because of the
rapid spatial variation in screening along the vacuum direction.\cite{Qiu2016,Attaccalite2011}
These issues have produced large discrepancies in literature
even for properties of pristine 2D materials.\cite{Qiu2016,Rasmussen2016}
As an example, converging GW calculations of pristine monolayer MoS$_2$ requires
at least 6000 bands, 25~\Angst vacuum spacing, and a $24\times 24\times 1$
$k$-point grid for Brillouin zone integration.\cite{Qiu2016}
Adopting such parameters for large supercell calculations containing defects
would make them impractical.

The slow convergence with respect to the number of bands can be overcome by using
a recent implementation of the GW method that does not explicitly require any
empty states as implemented in the WEST code,\cite{Pham2013,Nguyen2012,Ping2013, Rocca2014, Govoni2015,WESTcode}
based on density functional perturbation theory\cite{Baroni2001}
and the projective dielectric eigenpotential (PDEP) algorithm.\cite{Wilson2008,Wilson2009}

\comment{In this method based on density functional perturbation theory\cite{Baroni2001}
and the projective dielectric eigenpotential (PDEP) algorithm,\cite{Wilson2008,Wilson2009}
the GW equations are reformulated in terms of the projector over conduction states $\hat{P}_c$.
Using $\hat{P}_c = \hat{I} -\hat{P}_v$, where $\hat{I}$ is the identity
and $\hat{P}_v$ is the projector over the occupied states, this formalism
avoids any explicit reference to empty states and thereby avoids the associated
convergence issues present in traditional implementations.}

\begin{figure}
\includegraphics[width=\columnwidth]{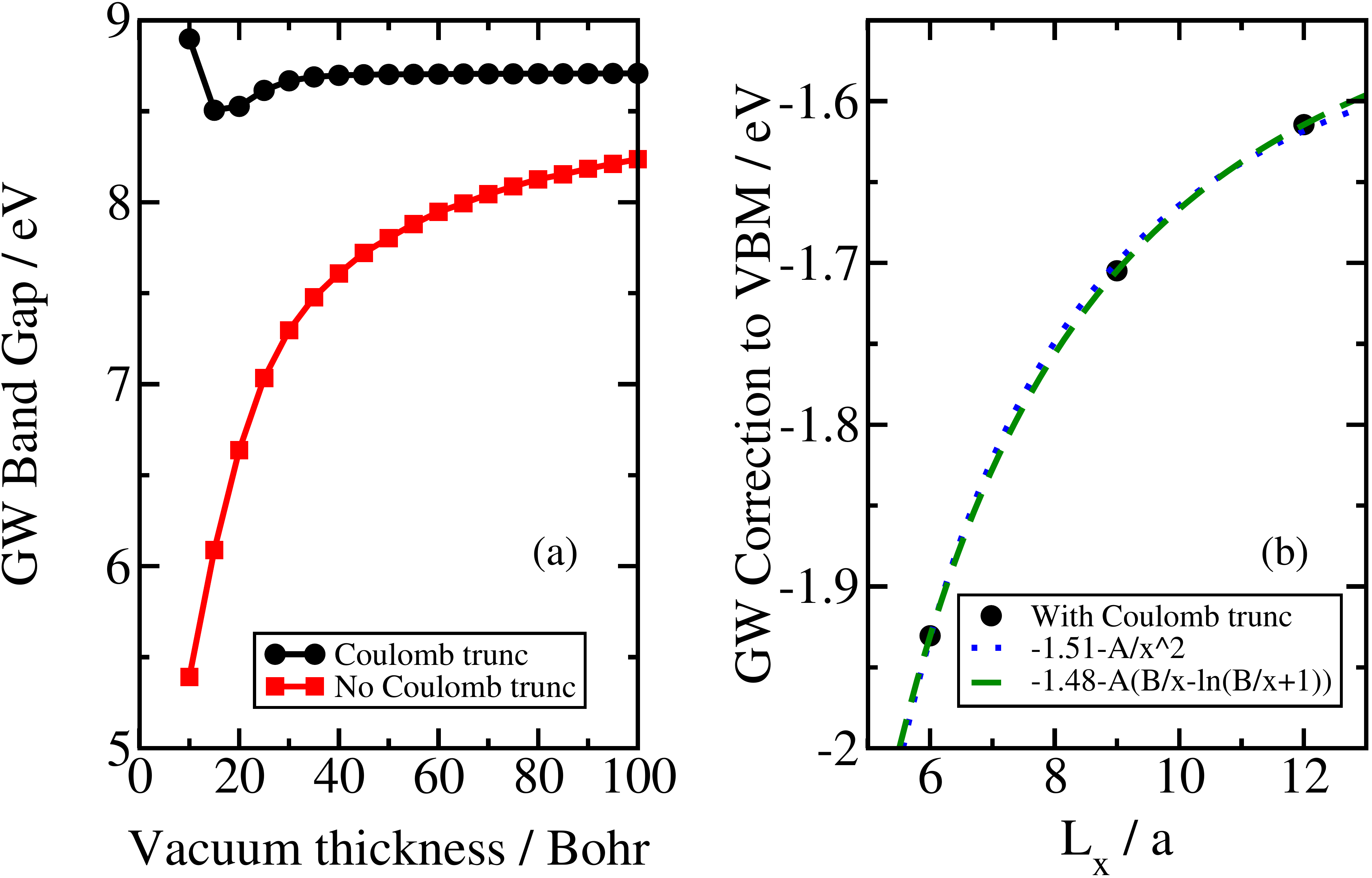}
\caption{(a) Coulomb truncation substantially improves
convergence of GW band gap of h-BN with respect to vacuum spacing 
(with a $3\times3$ a$^2$ lateral supercell size).
(b) GW correction to the VBM (valence band maximum) extrapolates reliably
with lateral supercell size. Black dots are the computed values with Coulomb truncation
and dot blue and dashed green lines are extrapolated values with two different formula.
\label{fig:gw-gap-compare-wo-coulombcutoff}}
\centering
\end{figure}

For 2D materials, the remaining convergence issues arise from the long-range nature of
the dielectric matrix and GW self-energy (in contrast to DFT), which have not been solved by current implementations.
Here the polarization of repeated images
in the direction perpendicular to the plane spuriously screens
the Coulomb interaction and lowers the QP gap.\cite{Qiu2016}
These image interactions can be avoided in the correlation part
of the self-energies by using a truncated Coulomb potential,
\begin{equation}
\bar{v}(\vb{k}) = \frac{4\pi}{k^2}\left(1-e^{-k_{xy}L_z/2}\cos\frac{k_zL_z}{2}\right),
\label{eq:Vtrunc}
\end{equation}
expressed here in reciprocal space~\cite{Ismail-Beigi2006}. In Eq.~(\ref{eq:Vtrunc}) we have
$\textbf{k}=\textbf{q}+\textbf{G}$, where $\textbf{q}$ is a wave vector in the first Brillouin zone
and $\textbf{G}$ denotes the reciprocal lattice vectors.
Fig.~\ref{fig:gw-gap-compare-wo-coulombcutoff}(a) shows that this truncation results in
excellent convergence with vacuum spacing for the GW QP gap of monolayer BN (Specifically, we performed G$_0$W$_0$ calculations
in which the self-energy is approximated from DFT states with the PBE exchange correlation functional~\cite{Perdew1996a}).
At 30~Bohr (16~\Angst), the QP gap is converged within 10 meV,
while the conventional treatment results in a smaller gap as
discussed above, which does not converge even at 100 Bohr.

When $\textbf{G}=0$ and $q_z=0$, the potential in Eq. (\ref{eq:Vtrunc}) diverges
as $2\pi L_z / q_{xy}$ for $q_{xy}\to 0$ and the inverse dielectric matrix
has a `dip' feature in this limit.\cite{Qiu2016}
Accordingly, around the $\Gamma$ point a fine $q$
mesh is required to compute absolute QP energies.\cite{Rasmussen2016}
Explicit $q$-mesh convergence is not practical for large supercell calculations with defects.
Instead, since discarding the $q_{xy}=0$ component introduces an error
proportional to $L^{-2}$, we extrapolate quasiparticle corrections to
the $L\to\infty$ limit from three supercell sizes ($L^2 = 6\times 6$,
$9\times 9$, $12\times 12$ $a^2$).
Fig.~\ref{fig:gw-gap-compare-wo-coulombcutoff}(b) shows that this extrapolation
works very well, with a deviation within 0.03~eV with respect to more sophisticated
models for the $\mathbf{q}\to 0$ contribution\cite{Rasmussen2016} (see note~\citenum{SCnote}).
The QP correction for the $12\times 12$ $a^2$ is converged within 0.1 eV compared to the extrapolated value.

We implemented the DFT charge correction scheme discussed above in JDFTx\cite{jdftx}
and the method to treat 2D materials in GW calculations in WEST.\cite{WESTcode}
Optimized geometry and DFT eigenvalues and wavefunctions are obtained
using Quantum ESPRESSO.\cite{Giannozzi2009} (See SI for further computational details.)

\begin{figure}
\begin{minipage}{0.45\columnwidth}
\includegraphics[width=\columnwidth]{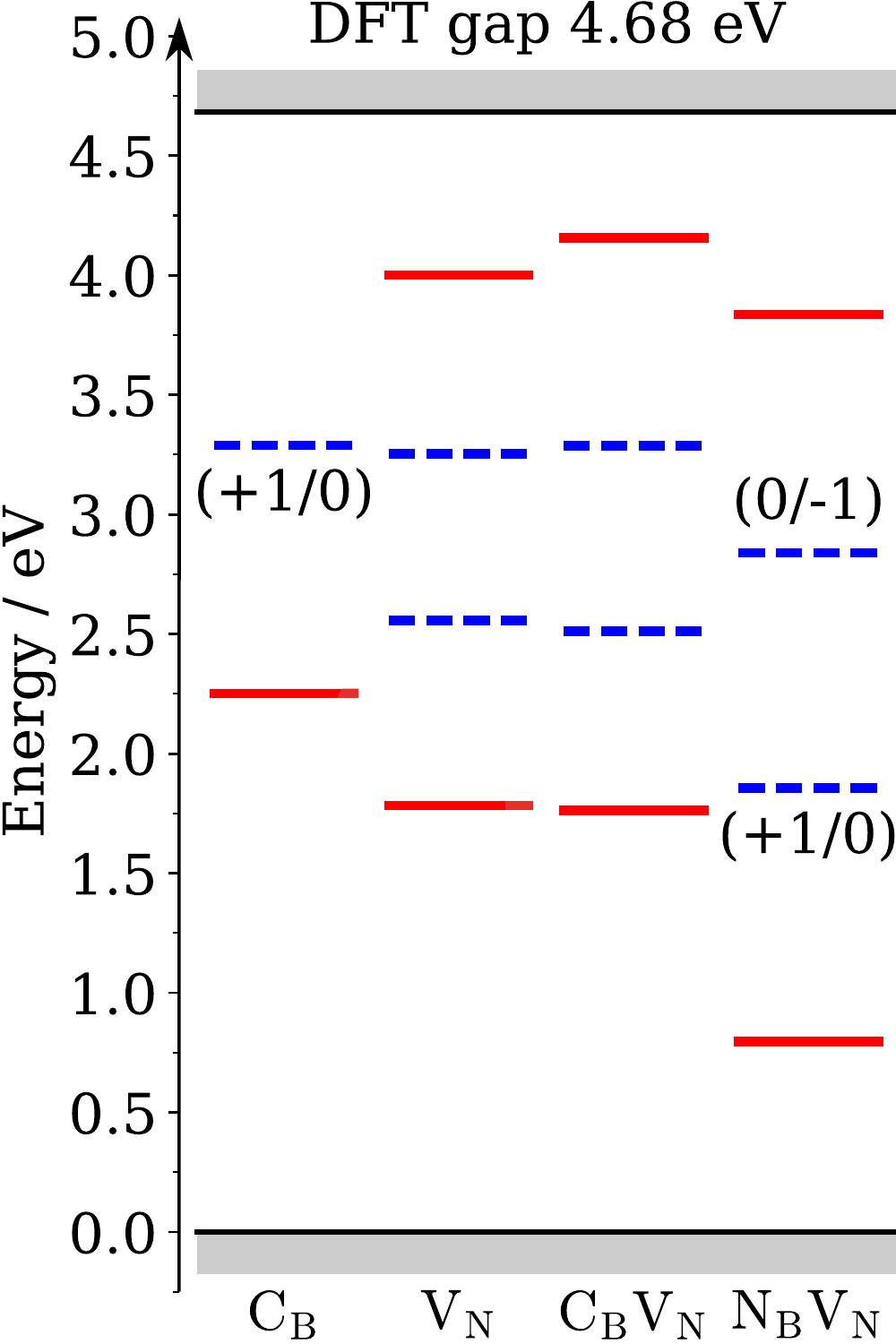}
\end{minipage}%
\begin{minipage}{0.45\columnwidth}
\includegraphics[width=\columnwidth]{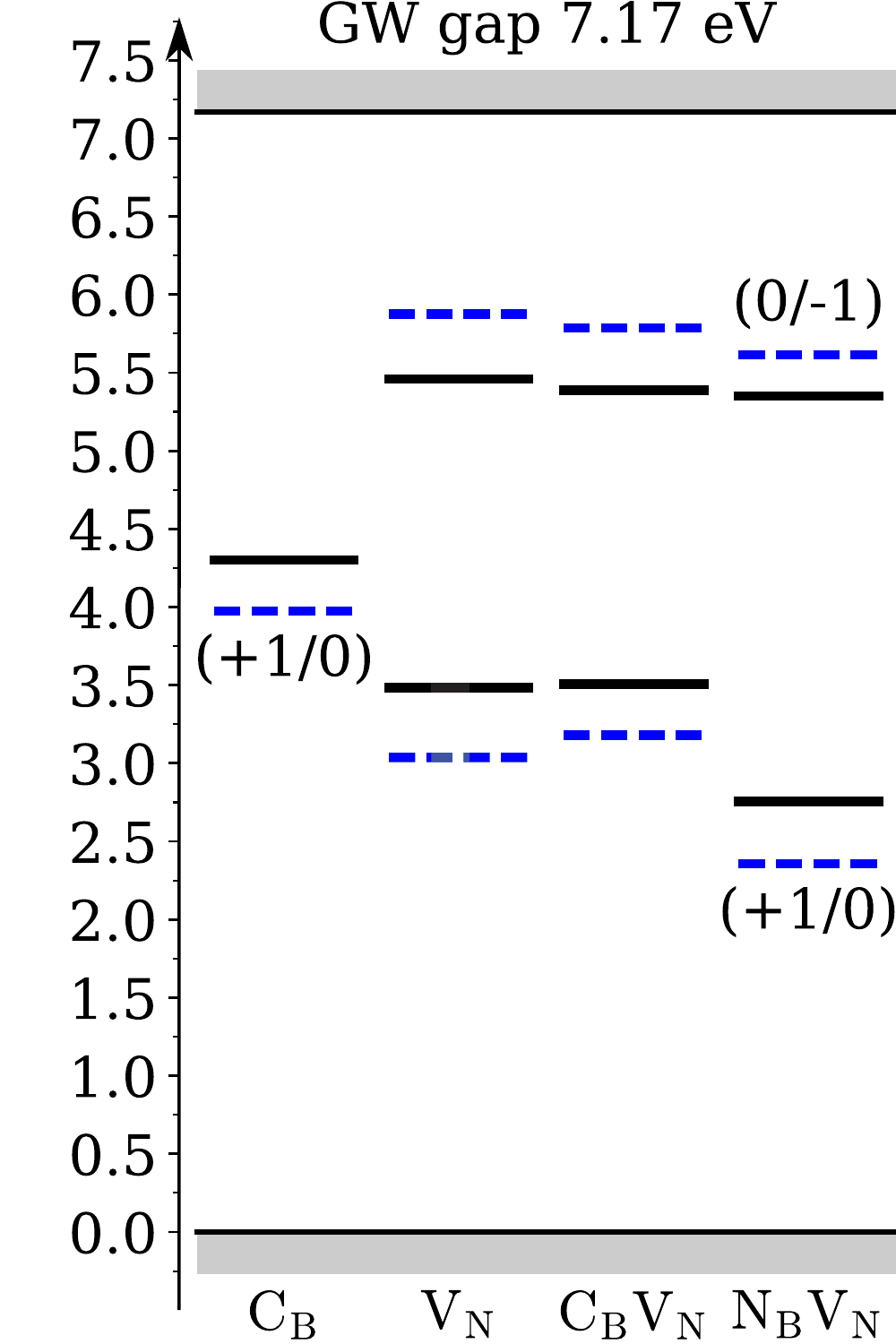}
\end{minipage}
\caption{Charge transition levels of various defects in h-BN computed
by DFT (left panel) and GW(right panel) methods.
Solid lines indicate thermodynamic charge transition levels
(Eq.~\ref{Ef} for DFT, Eq.~\ref{DFT+GW} for GW),
while dashed lines indicate optical charge transition levels.
Fermi level $\varepsilon_F$ is set to VBM of pristine h-BN. All defects have (+1/0) and (0/-1) CTLs inside the band gap, except \CB that only has (+1/0).
\label{fig:charge-transition-all}}
\centering
\end{figure}

Having eliminated all the roadblocks in calculating CTLs of 2D materials,
we now predict properties of the  simple (\CB (carbon substitution of boron), \NV (nitrogen vacancy)) and complex
(\CBNV, \NBNV (nitrogen substitution of boron adjacent to a nitrogen vacancy)) charged defects (\CBNV see Fig.~\ref{fig:opt-CBNV}).
As discussed earlier, promising candidate defects should have
stable high-spin states, localized and deep defect levels,
spin conserved excitations and anisotropic optical response.\cite{Jelezko2004,Weber2010}
Fig.\ref{fig:charge-transition-all} shows their optical
(without geometry relaxation, dashed lines) and
thermodynamic CTLs (solid lines) at both the DFT (left panel)
and GW (right panel) levels of theory (see note~\citenum{GWnote}).

Fig.\ref{fig:charge-transition-all} directly leads to several important conclusions.
At both DFT and GW levels of theory, all four type defects have deep CTLs
and localized defect wavefunctions (not shown). We note that
we also performed hybrid functional calculations
for the defective system, which partially correct the self-interaction errors in DFT,
and found the defect geometry, ground spin state and defect wavefunctions
are similar between hybrid and semilocal functionals (see SI for more details).
The difference of thermodynamic CTL by DFT+GW in Eq.~\ref{DFT+GW} and optical CTL by GW QP energies,
which is the geometric relaxation energy, is less than 0.5 eV.
The large difference between thermodynamic and optical CTLs in DFT
is consistent with the fact that the total energies in DFT are more reliable than eigenvalues,
as the optical CTLs are computed from eigenvalues directly,
which do not have physical meaning in DFT and can not be interpreted as the quasiparticle excitation energies.
In fact, correcting the VBM (and CBM) reference in the DFT thermodynamic CTLs
(using Eq.~\ref{Ef}) with GW QP energies, yields 0.1 eV difference
compared to the full DFT+GW calculations of the CTLs (using Eq.~\ref{DFT+GW}).

\begin{table}
\centering
\caption{Physical properties of defects in monolayer h-BN relevant for quantum technologies.
Below, ``S'', ``D'' and ``T'' denote singlet, doublet and triplet spin states respectively.
\label{tab:defect-property}}
\begin{tabular}{lllll}
	\hline\hline
Defects & \CB & \NV & \NBNV & \CBNV \\
\hline
Deep level & Yes & Yes & Yes & Yes \\
Spin at $q=0$ & D & D & D & T \\
Spin at $q=\pm 1$ & S & S & S & D\\[4pt]
\makecell[l]{Bright transition \\ between \\ defect states} & No & No &Yes & Yes\\[20pt]
\makecell[l]{Optical \\ anisotropy} & No & No & Yes & Yes \\
\hline\hline
\end{tabular}
\end{table}

All four defects have deep CTLs with the neutral state being stable for a wide range of $\varepsilon_F$,
but their spin and optical properties are rather different as Table~\ref{tab:defect-property} shows.
\CBNV center has a spin triplet ground state as shown in Fig.~\ref{fig:opt-CBNV} left panel, which is advantageous for
quantum applications,\cite{Weber2010} distinct from the doublet state in other defects.
Furthermore, we computed the optical transitions and absorption spectra for all defect cases and
found both \CBNV and \NBNV have bright defect-to-defect state transition
well separated by over 1 eV from any defect-bulk and bulk-bulk transitions.
A strong in-plane polarization anisotropy was also found in their absorption spectra (see Fig.~\ref{fig:opt-CBNV} and SI for details of absorption spectra and selection rules).

In summary, we developed a methodology to reliably calculate thermodynamic CTLs
in 2D materials by solving several critical issues in charged defect
formation energies and GW QP energies for 2D systems in general.
The source of difficulties originate from the highly anisotropic
and localized screening of 2D systems, which necessitates proper treatment
of electrostatic potentials of charges near a 2D plane and of the
screened Coulomb interaction in the GW approximation.
Using this methodology, we examined several possible defects in h-BN
and identified \CBNV center to be promising for quantum technologies, which has
multiple deep defect levels, a triplet ground state and bright defect-to-defect transitions.

\section*{Acknowledgment}
We thank Hosung Seo, Marco Govoni, Giulia Galli and Jairo Velasco for many useful discussions.
This research is supported by the startup funding of Department of Chemistry and Biochemistry at University of California, Santa Cruz,
as well as the Department of Materials Science and Engineering at Rensselaer Polytechnic Institute.
D.R. acknowledges financial support from Agence Nationale de la Recherche under grant number ANR-15-CE29-0003-01.
This research used resources of the Center for Functional Nanomaterials,
which is a U.S. DOE Office of Science Facility, at Brookhaven National Laboratory under Contract No. DE-SC0012704,
the National Energy Research Scientific Computing Center (NERSC),
a DOE Office of Science User Facility supported by the Office of Science
of the U.S. Department of Energy under Contract No. DEAC02-05CH11231,
and the Extreme Science and Engineering Discovery Environment (XSEDE),
which is supported by National Science Foundation grant number ACI-1548562.\cite{Supercomputer_XSEDE}

\section*{References}

%

\end{document}